# Deep-Learning-Accelerated Dopant Selection for High-k $HfO_2$ Dielectrics: A Disorder-Resolved Study of Y, Si and Al

Zunair Masroor[1], Bonwook Gu[1], Wonjoong Kim[1], Trinh Ngoc Le[1], Summal Zoha[1], Han-Bo-Ram Lee[1]*

[1]Department of Materials Science and Engineering, Incheon National University, Incheon 22012, South Korea

## Abstract

Hafnium oxide ($HfO_2$) is the cornerstone high-k dielectric in modern silicon technology. Since the constraints of silicon device fabrication rule out replacing the material itself, dopant incorporation is the principal means available to engineer its band gap ($E_g$) and dielectric constant ($\varepsilon$) within existing process flows. However, dopant selection is still largely empirical due to the coupled interplay among thermodynamic stability, electronic insulation, and dielectric response. Here, we present a high-throughput computational framework integrating special quasi-random structures (SQS), machine-learning potentials (SevenNet), and graph neural networks (ALIGNN) to systematically evaluate doped-$HfO_2$ compositions across three dopants (Al, Si, Y) and two technologically relevant polymorphs (monoclinic and orthorhombic). Our analysis uncovers a fundamental design principle: formation energy, band gap, and dielectric constant are decoupled parameters requiring application-specific prioritization rather than simultaneous optimization. Yttrium achieves the lowest formation energy (−3.763 eV/atom) and favors orthorhombic phase stabilization at process-compatible thermal budgets; silicon preserves near-pristine band gaps (~5.72 eV) critical for suppressing leakage in gate dielectric applications; and aluminum enables concentration-tunable band gap widening (5.6–5.9 eV) suited for voltage scaling. Validation against experimental literature and density functional theory (DFT) confirms quantitative accuracy (±0.02 eV band gap error for Si-doping, mean absolute error ≤ 0.001 eV/atom formation energy). This framework provides rational, property-decoupled guidance for dopant engineering in $HfO_2$-based dielectrics and related high-k oxide systems.

***Corresponding Authors**: hbrlee@inu.ac.kr

## 1. Introduction

Hafnium dioxide ($HfO_2$) is the cornerstone gate dielectric of modern silicon semiconductor technology, having replaced $SiO_2$ once direct-tunneling leakage made further $SiO_2$ scaling untenable.[1] With a dielectric constant of $k \approx 20–25$ and a wide band gap of ~5.7 eV, $HfO_2$ enables capacitance scaling at greater physical thickness while suppressing leakage, and its conformal deposition by atomic layer deposition (ALD) provides the sub-nanometer thickness control required for production.[2] More recently, the discovery of ferroelectricity in Si-doped nanoscale $HfO_2$ films has attracted significant interest for non-volatile memory applications.[3], [4]

A fundamental thermodynamic barrier nonetheless constrains $HfO_2$ performance: the bulk ground state is the non-polar monoclinic phase ($P2_1/c$), whose $k \approx 16$ falls well below that of the metastable higher-symmetry polymorphs.[5] Chemical doping is the primary strategy to suppress this phase, stabilize higher-k polymorphs, and simultaneously engineer the band gap.[6] Yet the dopants and the associated oxygen vacancies [7] that promote beneficial phase transitions also generate deep trap states and leakage pathways.[8] This establishes a fundamental dielectric–insulation trade-off that no single dopant resolves without application-specific prioritization.[3], [4], [5]

In ALD-grown films, dopant atoms are incorporated stochastically, producing local structural disorder that idealized ordered models cannot capture; small-cell approximations artificially enforce periodic dopant arrangements and yield unreliable predictions of phase stability and dielectric response.[10], [11], [12], [13] The special quasi-random structures (SQS) formalism addresses this by generating supercells that statistically reproduce the pair-correlation functions of a truly random alloy, closely mimicking chemically disordered ALD films.[13], [14], [15], [16] Structural fidelity, however, comes at a steep computational cost: capturing experimentally relevant dilute doping [17] without spurious periodic-image interactions requires supercells of several hundred atoms, which conventional DFT cannot relax across a broad dopant–phase landscape at tractable expense.[18] Prior studies have therefore compromised, either adopting the Virtual Crystal Approximation (VCA) or restricting scope to a single dopant species.[19], [20], [21]

Here we introduce AISHA (Atomistic Inference via SQS-driven High-throughput ALIGNN), a decoupled two-stage machine-learning workflow that resolves this bottleneck.[22], [23] In the first stage, SevenNet, an E(3)-equivariant machine-learning interatomic potential, relaxes large SQS supercells at near-DFT accuracy, making systems of several hundred atoms computationally accessible.[24], [25], [26], [27], [24], [28] In the second stage, the Atomistic Line Graph Neural Network (ALIGNN) evaluates the electronic structure of these relaxed configurations, using both bond lengths and bond angles to capture the local distortions that govern band gap evolution in doped oxides.[29], [30], [31] AISHA thus enables systematic high-throughput exploration of dopant chemistry, concentration, and phase stability in $HfO_2$.

We apply this framework to the two technologically dominant polymorphs, monoclinic and orthorhombic, generating 3×3×3 SQS supercells for three dopants commonly introduced during ALD processing: Si, Al, and Y.[9], [32] Across all dopant species, phases, and concentrations, we characterize band gap and dielectric constant, yielding a unified quantitative map of how dopant identity and concentration jointly govern Eg and k. Predictions are validated against DFT reference values from the Materials Project and experimental measurements from the literature, and the resulting database provides a validated platform for rational high-k oxide engineering and ALD process design.

## 2. Results & Discussion

All results presented in this section were generated using the AISHA framework: 3×3×3 SQS supercells (324 atoms, 108 Hf substitution sites) were relaxed with the SevenNet MLIP, and formation energy, MBJ band gap, and electronic dielectric constant were subsequently predicted using pre-trained ALIGNN models. Full computational details, including SQS generation parameters, relaxation convergence criteria, and model checkpoints, are provided in Section 4 (Computational Methodology).

The complete computational framework employed in this study is schematically illustrated in Fig. 1 and detailed as a step-by-step pipeline in Fig. 2. As shown in Fig. 1, the workflow proceeds through four sequential stages: (*i*) initialization of phase-specific supercells from experimentally and computationally validated $HfO_2$ polymorphs, (*ii*) generation of substitutionally disordered structures via the SQS algorithm, (*iii*) full geometric relaxation using a machine-learning interatomic potential (MLIP), and (*iv*) high-throughput property prediction via graph neural network (GNN) inference. Fig. 2 provides the corresponding flowchart of this pipeline, organized into three color-coded stages, SQS generation (ATAT mcsqs), structural relaxation (SevenNet MLIP), and property prediction (ALIGNN), explicitly illustrating the data and structural transformations at each stage from SQS generation through SevenNet-based relaxation to ALIGNN-based property extraction.

We retrieved the two thermodynamically relevant polymorphs of pristine $HfO_2$; monoclinic (m, space group $P2_1/c$) and orthorhombic (o, $Pca2_1$) from the Materials Project database, selecting the most thermodynamically stable entry for each phase (energy above the convex hull < 0.05 eV/atom) to ensure that subsequent doping calculations began from reliable reference geometries.[33] The monoclinic phase serves as the ambient ground-state structure, while the orthorhombic phase represents the ferroelectrically active metastable polymorph of primary relevance for high-k gate dielectric applications.[33] Each primitive cell contained the minimal atomic configuration consistent with $HfO_2$ stoichiometry, with cell parameters and atomic coordinates precisely defined by their crystallographic symmetry.

To eliminate the finite-size error inherent in small supercell calculations, whereby dopants in undersized cells artificially interact with their periodic images, generating spurious elastic and electrostatic contributions; each primitive cell was expanded to a 3×3×3 supercell, yielding approximately 324 atoms per structure corresponding to 108 available Hf substitution sites, such that each additional dopant atom represents an increment of approximately 0.9 at.% (Table S1). This expansion serves two essential purposes: (1) it sufficiently dilutes dopant-dopant interactions, enabling dopant concentrations spanning 2–20 at.% to be modeled without nearest-neighbor dopant-dopant adjacency that would artificially bias phase stability; and (2) it enlarges the configurational space available to the stochastic SQS algorithm, improving the statistical fidelity of the generated disorder models. The discrete mapping between each target concentration and the corresponding number of substituted Hf atoms is summarized in Table 1. The uniform supercell dimensions across both phases ensure direct, phase-by-phase comparability of dopant effects throughout the dataset.

For each supercell, substitutional doping was implemented using the SQS method, first introduced by Zunger et al.[34] and widely validated for disordered oxide systems including yttria-stabilized zirconia and gadolinia-doped ceria.[35] The SQS method solves a combinatorial optimization problem: given a target dopant concentration, dopant atoms are placed into the host supercell such that the radial pair and multi-site correlation functions of the resulting periodic structure statistically reproduce those of a true random alloy, minimizing artificial periodicity. Specifically, the optimization minimizes the sum of squared deviations between the target random-alloy correlations and the correlations computed from the trial configuration, explored via a Monte Carlo annealing algorithm.[14], [15] In this work, SQS structures were generated using the Monte Carlo-based mcsqs code within the Alloy Theoretic Automated Toolkit (ATAT), a widely adopted implementation that iteratively improves correlation matching through stochastic site-swap moves driven by a simulated-annealing schedule until convergence. As shown in SQS-generation stage (left column) of Fig. 2, this proceeds from the base $HfO_2$ supercell through generation of the ATAT input file (rndstr.in), which defines the parent lattice and the fractional site occupancies representing the target random alloy at each dopant concentration, followed by the Monte Carlo mcsqs search with soft-stop convergence, and parsing of the resulting best-fit structure file (bestsqs.out) into the final doped structure. The validity of SQS-based modeling for substitutional doping in $HfO_2$-related fluorite-structured oxides has been demonstrated in prior studies employing analogous workflows for isovalent and aliovalent impurities.[36]

The selection of Y, Al, and Si reflects their established use as ALD dopant precursors for $HfO_2$ films, where they are routinely incorporated to tune phase stability and dielectric response. The targeted concentration range of 2–20 at.% spans the experimentally accessible doping window achievable through ALD cycle ratio control, from dilute stabilization regimes to heavily doped compositions. The two polymorphs selected, monoclinic and orthorhombic, correspond directly to the phases observed in ALD-grown $HfO_2$ films under low and moderate thermal budgets respectively, making the computational dataset directly interpretable in terms of process outcomes. We generated SQS structures for three dopants; Yttrium (Y), Aluminum (Al), and Silicon (Si) at 10 distinct concentration values (2, 4, 6, 8, 10, 12, 14, 16, 18, and 20 at.%) across both $HfO_2$ polymorphs, yielding a total of 60 distinct SQS supercells (3 dopants × 2 phases × 10 concentrations), corresponding to the output shown at the base of SQS-generation stage in Fig. 2.

All three dopants are treated as substitutional impurities occupying $Hf^{4+}$ cation sites rather than interstitial positions. This treatment is well supported by both first-principles and experimental evidence: DFT studies of Al-, Y-, La-, and Si-doped $HfO_2$ consistently find that these dopants energetically prefer substitution at the Hf site, where they actively modulate the relative stability of the monoclinic, tetragonal, and polar orthorhombic phases. [37], [38], [39], [40] Experimentally, ALD-grown yttrium- and other-doped $HfO_2$ films are similarly described in terms of cation substitution into the fluorite-derived lattice, consistent with the gate-dielectric and ferroelectric behavior observed. [39], [40] The interstitial configuration is both energetically unfavorable, owing to the limited interstitial volume in the dense fluorite-derived framework, and inconsistent with the observed phase-stabilization behavior, justifying the substitutional model adopted here.

For isovalent substituents the dopant valence matches $Hf^{4+}$ directly, whereas for aliovalent substituents (e.g., $Y^{3+}$ replacing $Hf^{4+}$) the resulting charge imbalance of −1 per substitution is expected to be compensated by oxygen vacancies in experimental systems. Within the present AISHA framework, however, formal charge-neutral substitution is assumed following the standard convention in high-throughput computational screening studies of aliovalent oxide dopants. [41] This assumption is consistent with the scope of the present work, which targets comparative trends in formation energy and electronic properties across the dopant-concentration-phase landscape rather than absolute defect thermodynamics.

The computationally demanding task of structural relaxation was performed using SevenNet, a scalable E(3)-equivariant MLIP trained on extensive DFT datasets, enabling near-DFT accuracy at a fraction of the computational cost.[26], [27], [28] Starting from each SQS structure, atomic positions and cell vectors were iteratively optimized using a variable-cell scheme to a force convergence criterion of $f_{max} < 0.01$ eV/Å (implemented via the ExpCellFilter and FIRE optimizer, as shown in the structural-relaxation stage of Fig. 2) and an energy convergence threshold of $10^{-6}$ eV. SevenNet achieves a mean absolute error (MAE) of ~0.001 eV/atom for energy and $R^2 > 0.998$ for force predictions on $HfO_2$-related systems, providing structural equilibrium geometries of sufficient accuracy to capture the relative thermodynamic stabilities of the doped polymorphs.[26], [27] The complete relaxation pipeline for all 60 structures is illustrated in the central stage of Fig. 2, which traces the flow from SQS-structure input and SevenNet-MF-ompa potential loading through ASE-based structure reading and calculator attachment to the final relaxed geometries.

To characterize the electronic and functional properties of the SevenNet-relaxed configurations, we employed ALIGNN; an atomistic line graph neural network for materials property prediction, to predict three classes of properties directly relevant to high-k gate dielectric performance: (1) thermodynamic stability, quantified by formation energy ($E_f$, eV/atom); (2) electronic insulation, quantified by the band gap computed under the modified Becke-Johnson (MBJ) exchange–correlation correction ($E_g$, eV); and (3) optical dielectric response, quantified by the electronic dielectric constant (ε). As shown in the property-prediction stage (right column) of Fig. 2, each relaxed structure is converted from its CIF representation through ASE to JARVIS interface and DGL multigraph construction, after which three pre-trained ALIGNN models perform inference and output the $E_f$, $E_g$ and ε to a consolidated CSV file. These three quantities collectively define the device suitability of doped $HfO_2$, addressing both thermodynamic phase stability and the electronic insulation requirements critical for leakage suppression in advanced CMOS gate stacks.

Thermodynamic stability governs whether a doped $HfO_2$ configuration is physically realizable and determines the energetic hierarchy among competing polymorphs. To quantify stability, the formation energy of each doped structure was computed as:

$$\Delta E_f = \frac{E_{\text{total}}(\text{Hf}_{1-x}M_x\text{O}_2) - (1-x)\mu(\text{HfO}_2) - x\,\mu(M)}{N_{\text{atoms}}} \quad (1)$$

where $E_{total}$ is the ALIGNN-predicted total energy of the doped supercell, $x$ is the substitutional dopant fraction, $\mu(HfO_2)$ is the chemical potential of the undoped host referenced to the monoclinic ground state, and $\mu$(M) is the chemical potential of dopant element M in its standard elemental reference state. Negative $\Delta E_f$ values confirm that the doped configuration is thermodynamically stable relative to phase separation into the undoped host and elemental dopant reservoir.

Fig. 3(a) illustrates the ALIGNN inference stage, in which a relaxed crystal structure is converted into an atomistic line graph representation to predict formation energy, band gap, and electronic dielectric constant. Before analyzing doped systems, the accuracy of the SevenNet + ALIGNN pipeline was benchmarked against established reference values for the two undoped $HfO_2$ polymorphs considered in this work. As shown in Fig. 3(c), the ALIGNN-predicted formation energies for pure monoclinic ($P2_1/c$) and orthorhombic ($Pca2_1$) $HfO_2$ are −3.751 eV/atom and −3.787 eV/atom, respectively, compared to the Materials Project GGA-PBE reference values of −3.617 eV/atom (monoclinic, mp-352) and −3.581 eV/atom (orthorhombic, mp-685097), each taken as a single most-stable computed entry rather than a range. The ALIGNN absolute values are systematically more negative than GGA-PBE, which is expected given that ALIGNN was trained on a diverse multi-functional dataset that captures beyond-GGA correlation effects; analogous systematic offsets between ML-GNN formation energies and standard DFT have been documented in the literature.[42] Critically, both frameworks reproduce an inter-polymorph energy difference of ~36 meV/atom, consistent with the established consensus that the orthorhombic phase is metastable at ambient conditions.[43] The minor inversion in relative ordering between ALIGNN and GGA-PBE does not affect concentration-dependent trends, as the 36 meV/atom inter-polymorph gap lies within the known resolution limit of both methods for closely competing $HfO_2$ phases. Literature star markers on Fig. 3 (d–f), drawn from Pavoni et al.[44], are reported in cohesive energy units per formula unit and are not directly numerically comparable in absolute value, but their concentration-dependent trends are almost consistent with our predictions.

Fig. 3(d–f) presents the phase-averaged ALIGNN-predicted formation energies for Y-, Al-, and Si-doped $HfO_2$ across 2–20 at.% in both polymorphs. All 60 configurations yield negative formation energies throughout, confirming thermodynamic feasibility of substitutional doping for all six dopant-phase combinations. Yttrium achieves the most favorable formation energies (monoclinic: −3.816 → −3.728 eV/atom; orthorhombic: −3.801 → −3.725 eV/atom), with a remarkably flat concentration dependence over 2–12 at.% (variation < 0.026 eV/atom), indicative of high solid solubility driven by the near-optimal ionic size match of $Y^{3+}$ (1.019 Å) to the $Hf^{4+}$ site (0.97 Å). These trends are in quantitative agreement with Materlik et al.[43], who reported negative enthalpy of mixing for Y at all studied concentrations. It should be noted that the annotated literature values represent individual reported data points rather than averages, and that the Y-doped formation energies drawn from experimental and oxide-referenced studies (≈ −8 eV/atom) are expressed on a different reference basis than the per-atom, elemental-reference convention adopted here and by the Materials Project. The resulting near-constant offset reflects this definitional difference rather than a discrepancy in the underlying physics; critically, both our predictions and the literature consistently yield negative formation energies, preserving the same thermodynamic conclusion of dopant solubility.

The Al- and Si-doped literature values, computed on a basis consistent with ours, fall within 0.2–0.6 eV/atom of the ALIGNN predictions, confirming quantitative reliability where a like-for-like comparison is possible. Aluminum displays competitive stability at low concentration (monoclinic: −3.802 eV/atom at 2 at.%) but degrades steadily with increasing content (Δ = 0.209 eV/atom to 20 at.%), driven by the size mismatch of $Al^{3+}$ (0.535 Å) generating progressive elastic strain. Materlik et al.[43] likewise found that Al substitution favors the tetragonal phase at high concentration and produces only field-induced rather than intrinsic ferroelectricity, consistent with the rapid thermodynamic deterioration observed here beyond ~14 at.%. Silicon exhibits a closely comparable formation-energy trend (monoclinic: −3.790 → −3.568 eV/atom; Δ = 0.222 eV/atom), with a near-linear degradation of similar magnitude to Al, reflecting continuous strain accumulation from the small $Si^{4+}$ radius. Künneth et al.[37] demonstrated that Si substitution is fundamentally less thermodynamically favorable than rare-earth dopants owing to the small $Si^{4+}$ radius (0.40 Å) and oxygen vacancy compensation requirements, while Falkowski et al.[45] confirmed through high-throughput DFT that Si-doped $HfO_2$ exhibits large inter-configuration energy variance from dopant-dopant interactions, consistent with the concentration-driven degradation in our dataset.

The literature star markers overlaid on Fig. 3(d–f) reveal a systematic offset between the ALIGNN-predicted formation energies and previously reported DFT values, with literature values consistently more negative by several eV/atom. This discrepancy arises from well-documented methodological non-equivalences, including differences in elemental reference state conventions, the choice of exchange-correlation functional (PBE, PBEsol, or HSE06), and corrections applied to $O_2$ binding energy, each of which can shift absolute formation energies by 1–3 eV/atom independently.[46], [47] Additionally, the ALIGNN model was trained on Materials Project GGA-PBE energetics, which systematically underestimates the stability of strongly ionic oxides relative to higher-level functionals.[48] Despite this absolute offset, the heatmaps confirm the internal consistency of our model: the phase-dependent stability ordering (monoclinic more stable than orthorhombic across all dopants and concentrations) and relative inter-dopant rankings remain physically meaningful within our computational framework. Cross-study comparison of absolute values is most appropriately made against the uniform Materials Project reference shown in Fig. 3(c).

To quantify polymorph competition, we evaluate the relative phase stability, defined as the formation-energy difference between the orthorhombic and monoclinic phases, a standard metric for assessing the competition in $HfO_2$-based systems: [44], [49], [50]

$$\Delta E = E_f(\text{ortho}) - E_f(\text{mono})(\text{eV/atom}) \qquad (2)$$

where $\Delta E > 0$ indicates monoclinic preference and $\Delta E < 0$ indicates orthorhombic stabilization. The concentration-resolved ΔE trajectories in Fig. 3(b) reveal a clear dopant-dependent divergence in phase selectivity. Y-doping is the only system that crosses the phase boundary: starting at +0.024 eV/atom at 2 at.%, ΔE decreases to near-degeneracy at 10 at.% (+0.001 eV/atom) and reaches an orthorhombic-preferred minimum at 14 at.% (−0.014 eV/atom), consistent with experimental reports that the ferroelectric orthorhombic phase is optimally stabilized in the 6–14 at.% Y window.[51] In contrast, Al and Si maintain strictly positive ΔE throughout (Al: +0.010 to +0.031 eV/atom; Si: +0.016 to +0.030 eV/atom), with the monoclinic phase thermodynamically preferred at all studied concentrations; consistent with Materlik et al.[43] and Künneth et al.[37], who attribute any orthorhombic promotion in Al- and Si-doped systems to kinetic crystallization mechanisms rather than intrinsic thermodynamic driving forces.

The $E_g$ is the primary determinant of a dielectric material's ability to suppress leakage current, for example, gate oxides require $E_g > 5$ eV to maintain adequate tunneling barriers at scaled thicknesses. We analyze the ALIGNN-MBJ band gap predictions for Y-, Al-, and Si-doped $HfO_2$ across both polymorphs and the full 2–20 at.% concentration range, examining phase-averaged trends and dopant-resolved concentration dependencies against Materials Project benchmarks and available literature data. A critical limitation of standard DFT is severe band gap underestimation; the Materials Project GGA-PBE references for undoped $HfO_2$ yield only 4.07 eV (monoclinic) and 4.39 eV (orthorhombic); far below the experimental consensus of ~5.7 eV.[52], [53] The ALIGNN models employed here were trained on the JARVIS-DFT database, which provides band gaps computed with the modified Becke-Johnson (MBJ) meta-GGA potential rather than GGA-PBE; this MBJ correction recovers the accuracy lost to standard DFT gap underestimation. As shown in Fig. 4(a), these predicted values fall within the experimental band gap range reported for $HfO_2$ thin films (~5.3 to 5.7 eV) [54], [55] confirming that the MBJ-trained ALIGNN model reproduces experimentally consistent gaps, in contrast to the GGA-PBE references that underestimate them by more than 1 eV. The phase-averaged ALIGNN-MBJ predictions across all three dopants yield 5.65–5.73 eV for the monoclinic phase and 5.74–5.88 eV for the orthorhombic phase consistently above the 5.6 eV threshold required for effective leakage suppression in gate dielectric applications[43]. Notably, the orthorhombic phase exhibits a systematically wider band gap than the monoclinic across all dopant and concentration combinations, a consequence of its polar $Pca2_1$ symmetry imposing a more asymmetric crystal field that lifts orbital degeneracies at the band edges. Our own ALIGNN baseline predictions for undoped $HfO_2$ (monoclinic: 4.58 eV; orthorhombic: 4.32 eV) fall near the GGA-PBE range, indicating that dopant incorporation itself contributes to band widening; an effect captured consistently by the ALIGNN-MBJ framework across all studied configurations.

Fig. 4(b–d) resolves the phase-specific MBJ band gap trajectories for each dopant. Y-doping produces the widest orthorhombic band gaps among the three dopants, peaking at 5.966 eV at 8 at.% before decreasing moderately to 5.764 eV at 20 at.%, consistent with the progressive breakdown of the polar lattice symmetry at high substitution levels. The Y-monoclinic band gap increases gently from 5.655 eV (2 at.%) to a plateau of ~5.769 eV (18 at.%), in good agreement with Pavoni et al.[44], [56], who reported HSE-calculated Y-monoclinic band gaps of 5.8 eV and 6.0 eV at 8% and 12% Y, respectively; our ALIGNN predictions (5.751 and 5.769 eV) underestimate these by ~0.1–0.3 eV, within the known systematic offset of ALIGNN-MBJ relative to hybrid HSE06 calculations. The orthorhombic Y predictions show a larger deviation from Pavoni et al.'s HSE values at high concentration (e.g., 5.850 vs. 6.8 eV at 16% Y), attributable to the well-known difficulty of static SQS supercells in capturing the long-range polarization ordering that HSE06 fully resolves.

Al-doping exhibits the flattest concentration dependence of the three dopants in the monoclinic phase (5.634–5.673 eV, variation < 0.04 eV), indicating excellent electronic stability across the full 2–20 at.% window, a desirable property for process-tolerant device fabrication. Orthorhombic Al band gaps decrease from 5.917 eV (2 at.%) to 5.693 eV (20 at.%), with literature star markers on Fig. 4(c) (5.53–5.93 eV from first-principles Al-doped $HfO_2$ studies) broadly overlapping the ALIGNN-predicted band, confirming quantitative reliability of the predictions in this system. Si-monoclinic phase shows a slight widening trend with concentration, reaching 5.817 eV at 20 at.%, the highest monoclinic value among all dopants while orthorhombic Si decreases from 5.891 to 5.673 eV, mirroring the Al-orthorhombic behavior. The literature stars in Fig. 4(d), drawn from Si-doped $HfO_2$ DFT and spectroscopic studies (5.81–5.99 eV for orthorhombic at low concentrations), are in reasonable agreement with ALIGNN predictions at low-to-mid concentration, with increasing scatter beyond 10 at.% consistent with the larger configurational variance in high-concentration Si SQS structures noted by Falkowski et al.[45]

The apparent scatter across the literature star markers in Fig. 4(b–d) reflects a physically meaningful distinction rather than a prediction failure. To aid interpretation, DFT and experimental literature values are plotted as separate star colors. When DFT literature values are compared against ALIGNN predictions within the same computational framework, agreement is within 0.003–0.133 eV across all dopant-phase combinations, well within the expected accuracy of the MBJ functional. Experimental values introduce additional spread attributable to film-specific factors inherent to ALD-grown samples, including deposition temperature, precursor chemistry, oxygen vacancy concentration, and the measurement technique employed (spectroscopic ellipsometry versus electron energy-loss spectroscopy), none of which are captured in a ground-state DFT calculation on an idealized SQS supercell. The two most displaced experimental points, Y-orthorhombic at 6.8 eV and Si-monoclinic at 5.21 eV, originate from studies with non-standard film stoichiometries, and their deviation from the ALIGNN predictions reflects configurational sensitivity at those conditions rather than a systematic model error. Across all three dopants, ALIGNN-MBJ predictions maintain band gaps comfortably above 5.6 eV in both phases throughout the studied concentration range, establishing the electronic feasibility of Y, Al, and Si doping for high-k gate dielectric applications.

The electronic dielectric constant (ε) governs the polarizability of the host lattice at optical frequencies and, together with $E_g$, defines the capacitive performance of a high-k gate dielectric. In this work we focus on the electronic (high-frequency) component ε, as it is the quantity directly predicted by the ALIGNN model used here and can be benchmarked consistently against both DFPT electronic permittivities and optical-frequency experimental measurements. The ionic (lattice) contribution, which dominates the static permittivity of $HfO_2$, requires explicit phonon or DFPT calculation of the Born effective charges and is therefore beyond the scope of the present ML-based screening; it represents a defined direction for future work. The electronic component nonetheless provides a meaningful and internally consistent basis for comparing the relative polarizability of the different dopant-phase systems, and it is directly linked to the refractive index through $n = \sqrt{\varepsilon}$. This section analyzes ALIGNN-predicted ε for Y-, Al-, and Si-doped $HfO_2$ in monoclinic and orthorhombic phases across 2–20 at.%, benchmarked against Materials Project DFPT references and available literature data.

The central finding of the dielectric analysis is a robust phase hierarchy that persists across all dopant systems and concentrations: the orthorhombic phase systematically delivers higher ε than the monoclinic phase, with phase-averaged means of 4.96 and 4.47, respectively, as shown in Fig. 5(a). This ~0.5-unit advantage originates from the non-centrosymmetric $Pca2_1$ symmetry of the orthorhombic phase, where off-center Hf cation displacements sustain an intrinsic spontaneous polarization that amplifies the ionic contribution to lattice polarizability.[57] A second universal trend is concentration-driven suppression: across all three dopants and both phases, ε decreases monotonically with increasing substitution as dopant atoms progressively disrupt the long-range coherent polarization response of the $HfO_2$ host lattice. These two trends together define the design space: high ε is achieved at low dopant concentration in the orthorhombic phase, while dielectric stability across a wide concentration window is a property of the monoclinic phase. The undoped ALIGNN baselines (monoclinic: 5.147; orthorhombic: 5.129) slightly exceed the Materials Project DFPT references (monoclinic: 4.81; orthorhombic: 5.01), consistent with the known tendency of ALIGNN to marginally overestimate ε relative to static DFPT for fluorite-structured oxides, and do not affect the validity of the concentration-dependent trends.

Fig. 5(b–d) reveals how each dopant navigates the phase hierarchy and concentration suppression identified above. Y-doping produces the highest orthorhombic ε at low concentration (5.334 at 2 at.%), which decreases overall to 4.772 at 20 at.%, consistent with the progressive disruption of the polar cation displacement pattern by the large $Y^{3+}$ ion (1.019 Å). Al-doping exhibits the strongest concentration-driven suppression in the monoclinic phase, where ε declines from 4.592 at 2 at.% to 3.878 at 20 at.% ($\Delta = 0.714$), the largest absolute change in the dataset, directly reflecting the intrinsically low polarizability of the Al–O bond progressively diluting the Hf–O lattice response.[43] Si-doping, by contrast, produces the most stable monoclinic ε profile (4.435–4.683, variation < 0.25), confirming the covalent rigidity of the Si–O bond as a stabilizing framework element. [58]

Validating these trends against literature requires separating DFT and experimental sources, as the two carry fundamentally different uncertainties. DFT literature values, where direct comparison within the same computational framework is possible, show good-to-excellent agreement with ALIGNN predictions: Al-monoclinic at 10–12 at.% agrees to within 0.010–0.020, Al-orthorhombic at 10 at.% to within 0.014, and Si-orthorhombic at 14–16 at.% to within 0.021–0.047. These near-quantitative matches confirm that the ALIGNN pipeline correctly captures the dielectric physics of substitutionally doped $HfO_2$ where benchmark DFT data exists. Experimental literature values show considerably larger scatter, with several measurements (Y-monoclinic: 2.03–2.27; Al-monoclinic at 18 at.%: 2.92; Si-monoclinic at 10 at.%: 3.53) falling 1–2.5 units below ALIGNN predictions. These low values are characteristic of ε measured by characterization technique on real materials with significant amorphous content, grain boundary scattering, or non-stoichiometric oxygen concentrations; conditions that substantially suppress the electronic dielectric response relative to a fully crystalline SQS supercell.[45] Rather than indicating model error, this systematic downward shift of experimental ε in defective films reinforces the interpretation that ALIGNN predictions represent an upper-bound estimate for defect-free crystalline configurations, against which real ALD films or other materials will fall depending on process conditions. Across all three dopants, the dielectric analysis establishes that orthorhombic-phase stabilization at low dopant concentration is the most effective route to maximizing ε, while monoclinic Si offers the most reproducible dielectric performance across the full 2–20 at.% processing window.

The preceding sections individually establish how formation energy, band gap, and electronic dielectric constant respond to dopant identity, crystal phase, and concentration across the Y, Al, and Si systems. This section synthesizes those trends into a unified design framework and, critically, situates the ALIGNN predictions within the context of experimentally established dopant performance hierarchies reported in the ALD literature. Fig. 6 consolidates the three preceding property analyses into a Robertson high-k trade-off map,[2] plotting concentration-averaged ε against MBJ band gap for each dopant-phase combination, with ±1σ bars reflecting spread across 2–20 at.%. Each data point is rendered as its dopant element symbol (Al, Si, Y), with upright text denoting the monoclinic phase and italic text the orthorhombic phase, while the symbol color encodes the concentration-averaged formation energy on a green scale in which darker shades indicate greater thermodynamic stability. This encoding allows all three design criteria, dielectric response, band gap, and phase stability, to be read simultaneously from a single map. The orthorhombic phase occupies the upper-right region across all three dopants, confirming that $Pca2_1$ symmetry simultaneously enhances polarizability and electronic insulation relative to the monoclinic ground state. This phase superiority is consistent with experimental studies, which have consistently reported that orthorhombic-stabilizing dopants, particularly Y, yield the most favorable combined dielectric and band gap performance among the common $HfO_2$ dopants.[44], [56] The ALIGNN predictions reproduce this hierarchy quantitatively: Y-orthorhombic achieves the highest band gap ($E_g = 5.91 \pm 0.07$ eV) and the strongest thermodynamic stability ($E_f = -3.777$ eV/atom) of all 3 dopant-phase combinations, alongside a dielectric constant ($\varepsilon = 4.96 \pm 0.19$) statistically indistinguishable from the Si-orthorhombic maximum ($4.998 \pm 0.25$). Y is therefore the only dopant that simultaneously leads in electronic insulation, thermodynamic driving force, and competitive dielectric response; a multi-property advantage that experimental observations have qualitatively captured but that the present framework quantifies across the full 2–20 at.% landscape for the first time.

The Robertson map also reveals important process-tolerance distinctions absent from single-property analyses. Al-orthorhombic and Si-orthorhombic are competitive alternatives to Y-orthorhombic at low concentration (2–8 at.%), where all three orthorhombic ε values converge near 5.1–5.2, but both show significantly larger concentration sensitivity (quantified by the standard deviation $\sigma$ of $\varepsilon$ across the 2–20 at.% range; Al-orthorhombic = 0.295, Si-orthorhombic = 0.251) than Y-orthorhombic ($\sigma$ = 0.187), imposing tighter experimental process control requirements to maintain target properties. The formation-energy color scale in Fig. 6 further separates these alternatives on stability grounds: Si-orthorhombic, despite reaching the highest dielectric constant in the dataset, is the least thermodynamically stable configuration ($E_f = -3.658$ eV/atom) and therefore appears in the palest shade, whereas Y-orthorhombic combines its leading band gap with markedly stronger stability. Notably, dielectric response and formation energy are essentially uncorrelated across the six centroids, confirming that high permittivity carries no intrinsic stability penalty or benefit and must be optimized as an independent design axis. Among monoclinic configurations, Y-monoclinic and Si-monoclinic exhibit the tightest dielectric stability across concentration ($\sigma$ = 0.066 and 0.072, respectively), making them the most reproducible options when the monoclinic phase is process-determined. Y-monoclinic is additionally the most thermodynamically stable configuration overall ($E_f = -3.784$ eV/atom), reinforcing yttrium as the most robust dopant choice irrespective of which phase the process ultimately selects. Al-monoclinic, despite delivering the flattest band gap profile in the entire dataset (5.633–5.674 eV, variation < 0.04 eV), records the weakest dielectric response ($\varepsilon = 4.21 \pm 0.26$), reflecting the intrinsically low polarizability of the Al–O bond and ruling it out for applications where dielectric enhancement is the primary objective.

The physical self-consistency of the predicted $\varepsilon$ values underpinning these recommendations is independently verified in Fig. S1, which plots the refractive index n derived from the Maxwell relation $n = \sqrt{\varepsilon}$ against the ALIGNN-predicted $\varepsilon$ for all dopant-phase-concentration combinations. The near-perfect adherence of all data points to the theoretical Cauchy curve ($R_2 \approx 1.000$) confirms that the predicted dielectric values satisfy classical electromagnetic dispersion theory across the entire compositional landscape, including at high dopant concentrations where configurational disorder is greatest. That no single dopant-phase combination is simultaneously optimal across all three properties confirms that high-k $HfO_2$ design cannot be reduced to single-metric screening, and that the AISHA framework demonstrated here provides an efficient, physically grounded pathway for navigating this multi-dimensional optimization space.

## 3. Conclusion

This study presents a high-throughput computational framework combining SevenNet machine-learning interatomic potential relaxation and ALIGNN graph neural network inference to systematically characterize the thermodynamic stability, electronic band gap, and optical dielectric constant of Y-, Al-, and Si-doped $HfO_2$ across monoclinic and orthorhombic polymorphs at 2–20 at.% dopant concentration. Across all 60 screened configurations, negative formation energies confirm the thermodynamic feasibility of substitutional doping for all dopant-phase combinations, with Y-doping uniquely capable of inducing thermodynamic phase switching toward the ferroelectrically active orthorhombic polymorph near 14 at.%, a crossover unattainable through the thermodynamic driving forces of Al or Si alone. ALIGNN-MBJ predictions confirm that all configurations maintain $E_g > 5.6$ eV, satisfying the electronic insulation requirement for advanced gate dielectric applications, while the orthorhombic phase consistently delivers superior dielectric response ($\varepsilon \approx 4.84$–$5.00$) relative to its monoclinic counterpart across all three dopants. The physical self-consistency of the predicted dielectric values is confirmed by near-perfect adherence to the Maxwell relation $n = \sqrt{\varepsilon}$ across all dopant-phase-concentration combinations ($R^2 \approx 1.000$), independently validating the ALIGNN predictions against classical electromagnetic dispersion theory. Multi-property analysis establishes that no single dopant-phase combination is universally optimal, but yields a clear application-specific hierarchy: Y-orthorhombic at 6–10 at.% delivers the best simultaneous thermodynamic stability, band gap, and dielectric response and is the overall recommended configuration for high-k gate dielectric applications; Si-orthorhombic at 2–8 at.% offers the highest dielectric response with moderate concentration tolerance; and Al-monoclinic provides the most uniform band gap across the full doping window for leakage-critical designs. Beyond the specific findings for $HfO_2$, this work demonstrates that ML-accelerated high-throughput screening at the dopant-phase-concentration resolution required for rational dielectric engineering is computationally tractable, providing an extensible framework for future studies of co-doping, oxygen vacancy engineering, and interfacial effects in next-generation high-k semiconductor systems.

## 4. Computational Methodology

### 4.1 Framework Overview

The complete computational workflow is illustrated in Figure 1, which outlines the four sequential stages of the high-throughput pipeline. In the Initialization stage, experimentally validated $HfO_2$ base structures are retrieved from the Materials Project database and expanded into large periodic supercells suitable for dilute doping. In the Generation stage, chemically disordered doped configurations are constructed using the SQS formalism to faithfully represent the stochastic nature of experimental doping. In the Relaxation stage, all generated structures undergo full variable-cell geometry optimization using the SevenNet MLIP, converging both atomic positions and cell parameters at near-DFT accuracy. Finally, in the Prediction stage, the relaxed structures are fed into the ALIGNN to extract key electronic and dielectric properties; band gap and electronic dielectric constant across all dopant–phase–concentration combinations. Together, these stages form a fully automated, end-to-end pipeline capable of systematically screening the doped $HfO_2$ design space without the prohibitive computational cost of direct DFT.

**4.2 Base Structure Retrieval and Supercell Construction**

Pristine $HfO_2$ base structures for the monoclinic (space group $P2_1/c$) and orthorhombic (space group $Pca2_1$) phases were retrieved from the Materials Project database using the mp-api Python client interfaced with pymatgen. For each target phase, a phase-filtered query was executed specifying the formula $HfO_2$ and the respective crystal system, with an initial energy-above-hull filter of 0–0.05 eV/atom to restrict retrieval to thermodynamically competitive polymorphs. Among returned entries, the structure with the lowest energy above the convex hull was selected as the reference configuration to ensure maximum thermodynamic representativeness. Retrieved structures were saved in CIF format with filenames encoding the phase label and Materials Project identifier to enable unambiguous traceability throughout the pipeline.

To achieve the supercell sizes necessary for representing dilute doping concentrations without spurious periodic image interactions, each base structure was expanded into a 3×3×3 supercell multiplying all lattice vectors uniformly. This expansion yielded supercells containing 324 atoms for the monoclinic phase (108 Hf + 216 O) and 324 atoms for the orthorhombic phase, providing a sufficient number of Hf substitution sites (108 per supercell) to span the target doping range of 2–20 at.% in 2% increments corresponding to 2 to 22 substituted Hf atoms, without the artificial dopant–dopant interactions that afflict smaller cells.

**4.3 Special Quasirandom Structure (SQS) Generation**

Chemically disordered doped configurations were generated using the SQS formalism as implemented in the mcsqs algorithm within the ATAT. SQS generation was performed by substituting Hf sites with three dopant species; Al, Si, and Y across ten doping concentrations (2, 4, 6, 8, 10, 12, 14, 16, 18, and 20 at.%) for each of the two target phases, yielding a total of 60 distinct doped supercell configurations.

For each dopant-concentration combination, a rndstr.in input file was constructed programmatically, encoding the lattice parameters, fractional atomic coordinates, and site occupancies of the 3×3×3 supercell. Hf sites were assigned mixed occupancies of the form $Hf_{1-x}D_x$, where D denotes the dopant species and x is the target molar fraction, while all O sites retained their native identity. Pair- and triplet-correlation cluster functions were pre-generated using mcsqs -2=6 -3=4, specifying a pair cutoff radius of 6 Å and a triplet cutoff of 4 Å, defining the correlation targets that the Monte Carlo search must reproduce to qualify as a valid SQS. The Monte Carlo optimization was then launched with the full supercell size specified via -n={total_atoms}, using a positional randomization seed (-ip=1) and a cell-shape weight (-wr=1) to balance structural diversity against lattice compatibility.

To ensure pipeline robustness during high-throughput execution, a soft-stop mechanism was implemented: if the mcsqs process did not converge naturally within a configurable time limit (default 1800 seconds per structure), a stopsqs sentinel file was written to the working directory, instructing ATAT to terminate gracefully and write out the best solution found up to that point. The best-candidate structure was then extracted from the bestsqs.out output file which encodes the SQS supercell with the lowest residual pair-correlation deviation from the ideal random alloy and parsed into a pymatgen Structure object for downstream processing. This soft-stop strategy ensures that the pipeline always produces a physically meaningful output even for configurations where the Monte Carlo search does not reach full convergence within the allotted wall time.

**4.4 Variable-Cell Structural Relaxation via SevenNet**

All 60 SQS structures were subjected to full variable-cell geometry optimization using SevenNet, an E(3)-equivariant graph neural network interatomic potential trained on the extensive MatPES dataset. Specifically, the SevenNet-MF-ompa multi-fidelity model was employed, which incorporates training data at both PBE and meta-GGA $r^2$SCAN levels of theory, providing a favorable balance between computational efficiency and energy–force accuracy. The SevenNet calculator was interfaced with the Atomic Simulation Environment (ASE) via the SevenNetCalculator class.

Relaxations were performed using the Fast Inertial Relaxation Engine (FIRE) optimizer applied to an ExpCellFilter object, which simultaneously relaxes both the internal atomic positions and the full cell metric tensor (lattice vectors and angles) under the constraint of no artificial hydrostatic bias (hydrostatic_strain=False). This approach allows the lattice to respond naturally to the local chemical environment introduced by the dopant, capturing symmetry-lowering distortions and volume changes that are physically essential for accurate property prediction.

Convergence was declared when the maximum residual force on any atom fell below 0.01 eV/Å, with a maximum of 1000 FIRE steps permitted per structure. A time step of dt = 0.1 fs was used for the FIRE integration. Upon successful convergence, the relaxed structure was written to a CIF file with the filename reflecting the relaxed tag in place of sqs, preserving the dopant, phase, and concentration metadata in the filename for downstream parsing.

**4.5 Property Prediction via ALIGNN**

Properties of all relaxed structures were predicted using ALIGNN, a graph neural network architecture that explicitly encodes both bond lengths and bond angles through a dual-graph representation an atom-bond graph and a bond-angle line graph, enabling the model to capture the subtle geometric distortions that govern property evolution in doped oxides. Three pre-trained ALIGNN model checkpoints from the JARVIS-ML model repository were employed:

- jv_mbj_bandgap_alignn: Band gap predicted at the modified Becke–Johnson (MBJ) level, providing accuracy comparable to hybrid functionals
- jv_formation_energy_peratom_alignn: Formation energy per atom (eV/atom), quantifying thermodynamic stability
- jv_epsx_alignn: Electronic contribution to the dielectric constant, directly governing high-frequency permittivity

Each relaxed CIF file was read via ASE and converted to a JARVIS Atoms object through a coordinate transformation that maps Cartesian positions to fractional coordinates using the inverse of the lattice matrix. The atomic graph and line graph were constructed using Graph.atom_dgl_multigraph, and the lattice tensor was passed as an additional input to preserve cell-level structural information. All models were executed in inference mode (torch.no_grad()) on GPU, with predicted values for physical properties bounded to non-negative values by taking the absolute value for band gaps, formation energy, and dielectric constants. The electronic refractive index was derived as $n = \sqrt{\varepsilon_\infty}$ from the predicted electronic dielectric constant. Results for all structures were aggregated into a master CSV file indexed by dopant species, phase, and doping concentration, with per-dopant CSV files additionally generated for targeted analysis.

## 5. Code availability

All custom Python scripts developed for this study including base structure retrieval, 3×3×3 supercell construction, ATAT-based SQS generation with soft-stop control, SevenNet variable-cell relaxation, and ALIGNN multi-property prediction are openly available in a public repository to ensure full reproducibility of the reported results. The complete codebase, along with usage instructions and example input files, can be accessed at: https://github.com/hbrl-research-group/SQS-Simulation.

## Figures

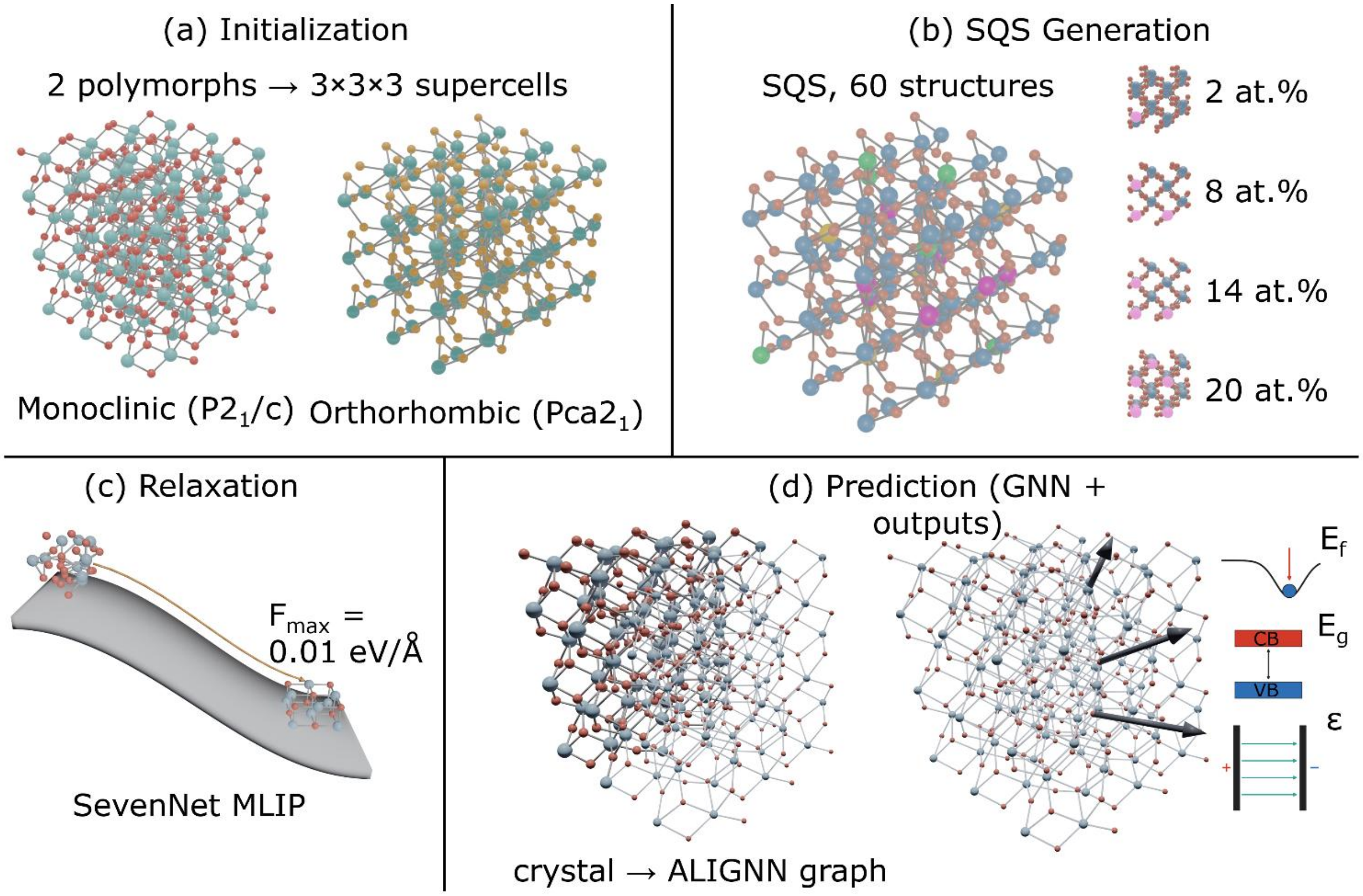

**Figure 1.** High-throughput computational pipeline for property prediction of doped $HfO_2$. (a) Initialization: Generation of 3×3×3 supercells from the pristine monoclinic ($P2_1/c$) and orthorhombic ($Pca2_1$) polymorphs of $HfO_2$. (b) SQS Generation: Introduction of Y, Al, and Si dopants at Hf sites using the Special Quasi-random Structure (SQS) methodology. A total of 60 distinct configurations are generated across ten dopant concentrations spanning 2 to 20 at.% (representative cells at 2, 8, 14, and 20 at.% are shown). (c) Relaxation: Structural relaxation and energy minimization driven by the SevenNet Machine Learning Interatomic Potential (MLIP), optimized to a force-convergence criterion of $F_{max} = 0.01$ eV/Å. (d) Prediction: Conversion of the relaxed structures into graph representations for processing by the Atomistic Line Graph Neural Network (ALIGNN), which predicts three core properties: formation energy ($E_f$), band gap ($E_g$), and electronic dielectric constant ($\varepsilon$).

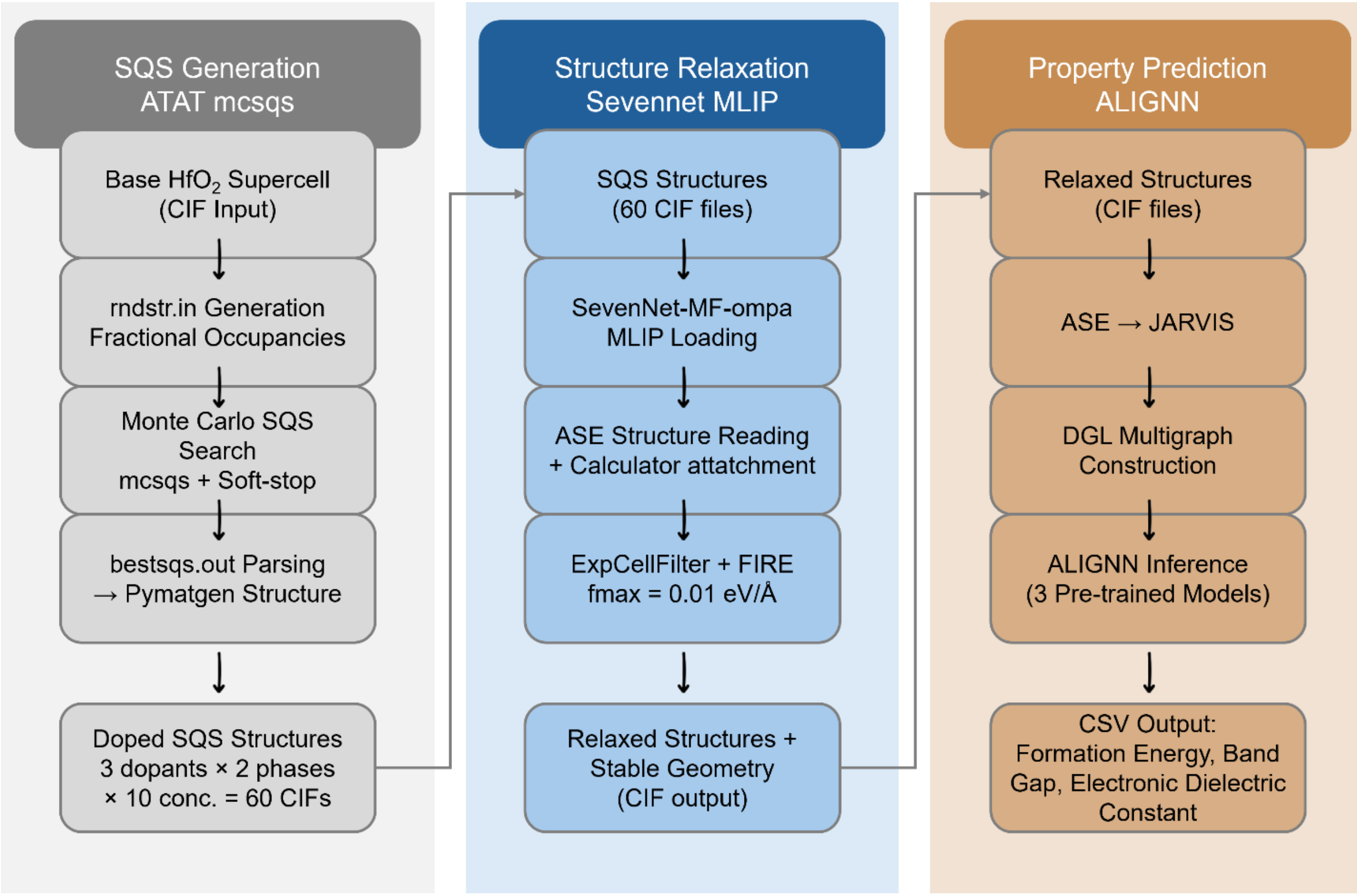

**Figure 2.** High-throughput computational pipeline for property prediction of doped $HfO_2$. *SQS Generation:* Monte Carlo sampling is used to generate 60 distinct Special Quasi-random Structures (SQS) spanning three candidate dopants across two host phases at concentrations from 2 to 20%. *Structural Relaxation:* Geometries undergo full variable-cell optimization via a machine-learning interatomic potential (SevenNet MLIP) to a strict force convergence ($f_{max}$= 0.01 eV/Å). *Property Prediction:* Graph neural networks (ALIGNN) evaluate the relaxed configurations to rapidly extract fundamental thermodynamic stability ($E_f$) and optoelectronic properties ($E_g$, ε). *Abbreviations:* ATAT, Alloy Theoretic Automated Toolkit; CIF, Crystallographic Information File; ASE, Atomic Simulation Environment; FIRE, Fast Inertial Relaxation Engine; JARVIS, Joint

Automated Repository for Various Integrated Simulations; DGL, Deep Graph Library; ALIGNN, Atomistic Line Graph Neural Network.

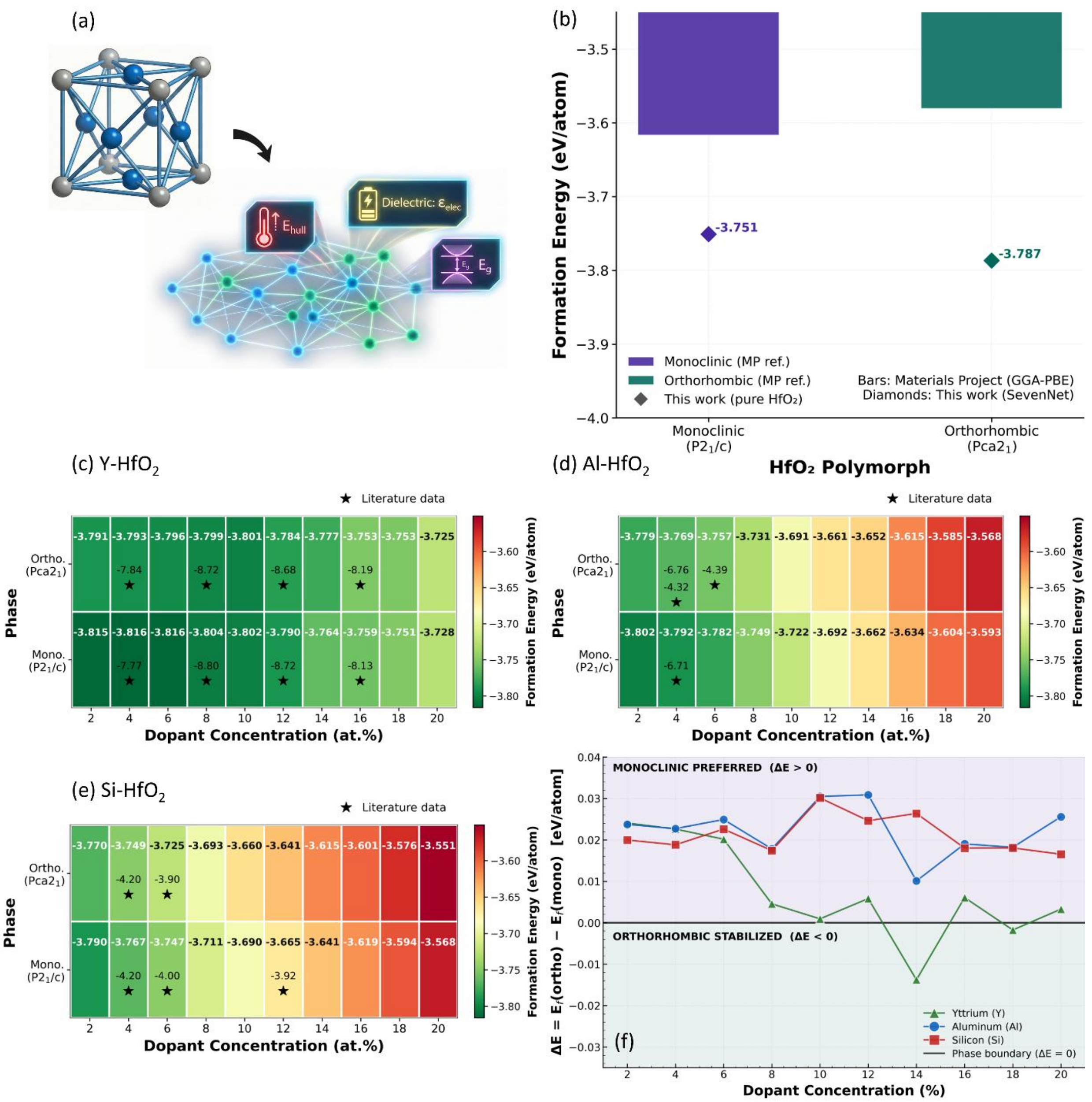

**Figure 3.** Thermodynamic stability of doped $HfO_2$. (a) Schematic of the ALIGNN inference stage, converting a relaxed crystal structure into an atomistic line graph to predict formation energy, band gap, and dielectric constant. (b) Formation energies of pure monoclinic and orthorhombic $HfO_2$, validating this work (diamonds) against Materials Project benchmarks (bars). (c–e) Formation

energy heatmaps versus dopant concentration (2–20 at.%) and phase for (c) Y-, (d) Al-, and (e) Si-doped $HfO_2$; green denotes greater stability (more negative $E_f$), with ALIGNN values annotated per cell and literature values (black stars) overlaid. (f) Phase preference map showing $\Delta E = E_f(\text{ortho}) - E_f(\text{mono})$ versus concentration; $\Delta E = 0$ separates the monoclinic-preferred ($\Delta E > 0$) from the orthorhombic-stabilized ($\Delta E < 0$) regime. All doped configurations exhibit negative formation energies, confirming thermodynamic feasibility throughout.

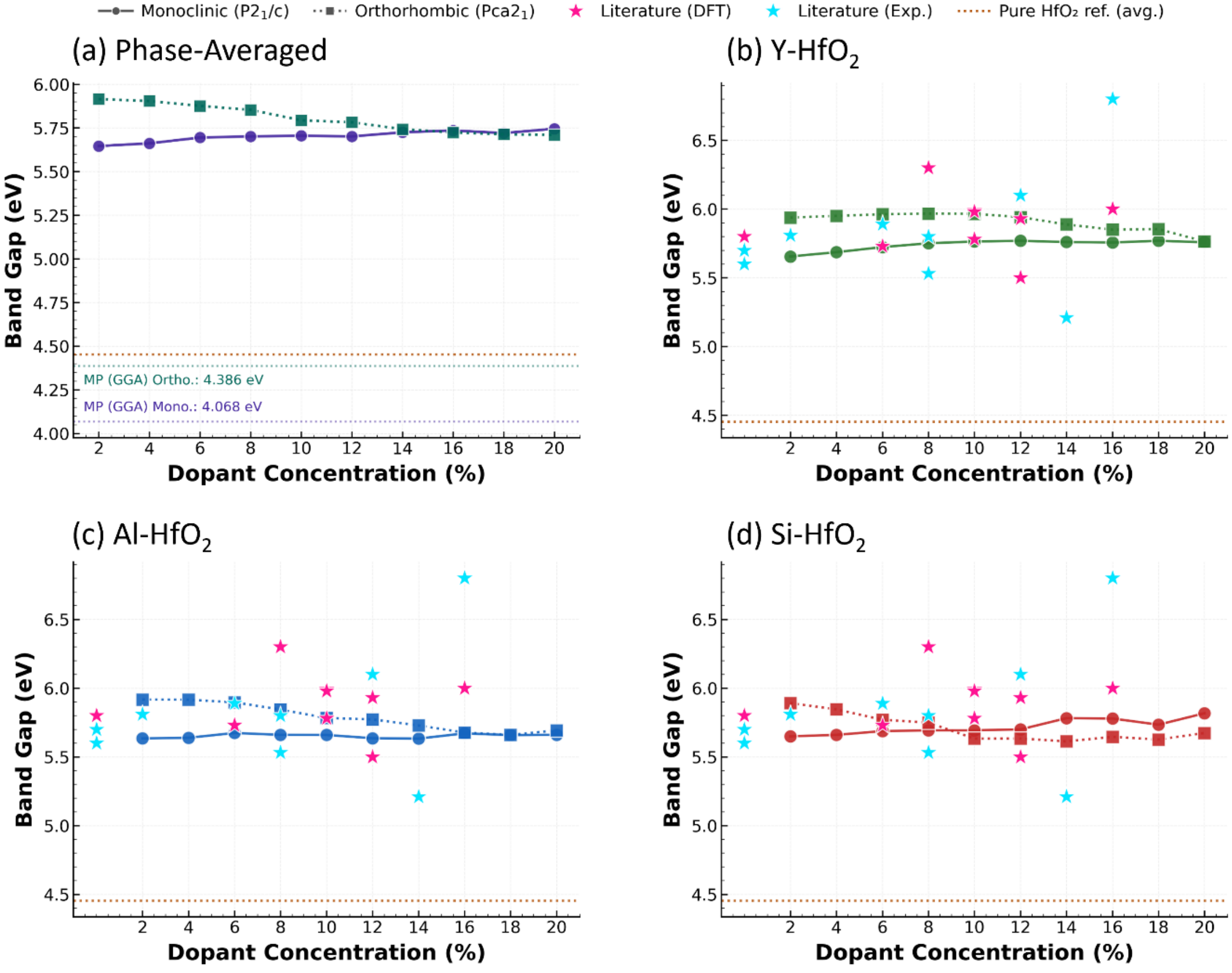

**Figure 4.** Band gap engineering in doped $HfO_2$. (a) Phase-averaged MBJ band gaps across all dopant systems for monoclinic (solid lines) and orthorhombic (dashed lines) phases, benchmarked against Materials Project GGA-PBE reference values (horizontal dashed lines). (b–d) Concentration- and phase-resolved band gaps for (b) Y-, (c) Al-, and (d) Si-doped $HfO_2$. ALIGNN predictions (lines with markers) are compared against DFT literature (pink stars) and experimental literature (turquoise stars), with the pure $HfO_2$ SevenNet baseline shown as a dotted reference line. Across all dopant systems, both phase selection and compositional tuning demonstrate the capacity to modulate $E_g$ toward the >5 eV threshold required for advanced high-k gate dielectrics.

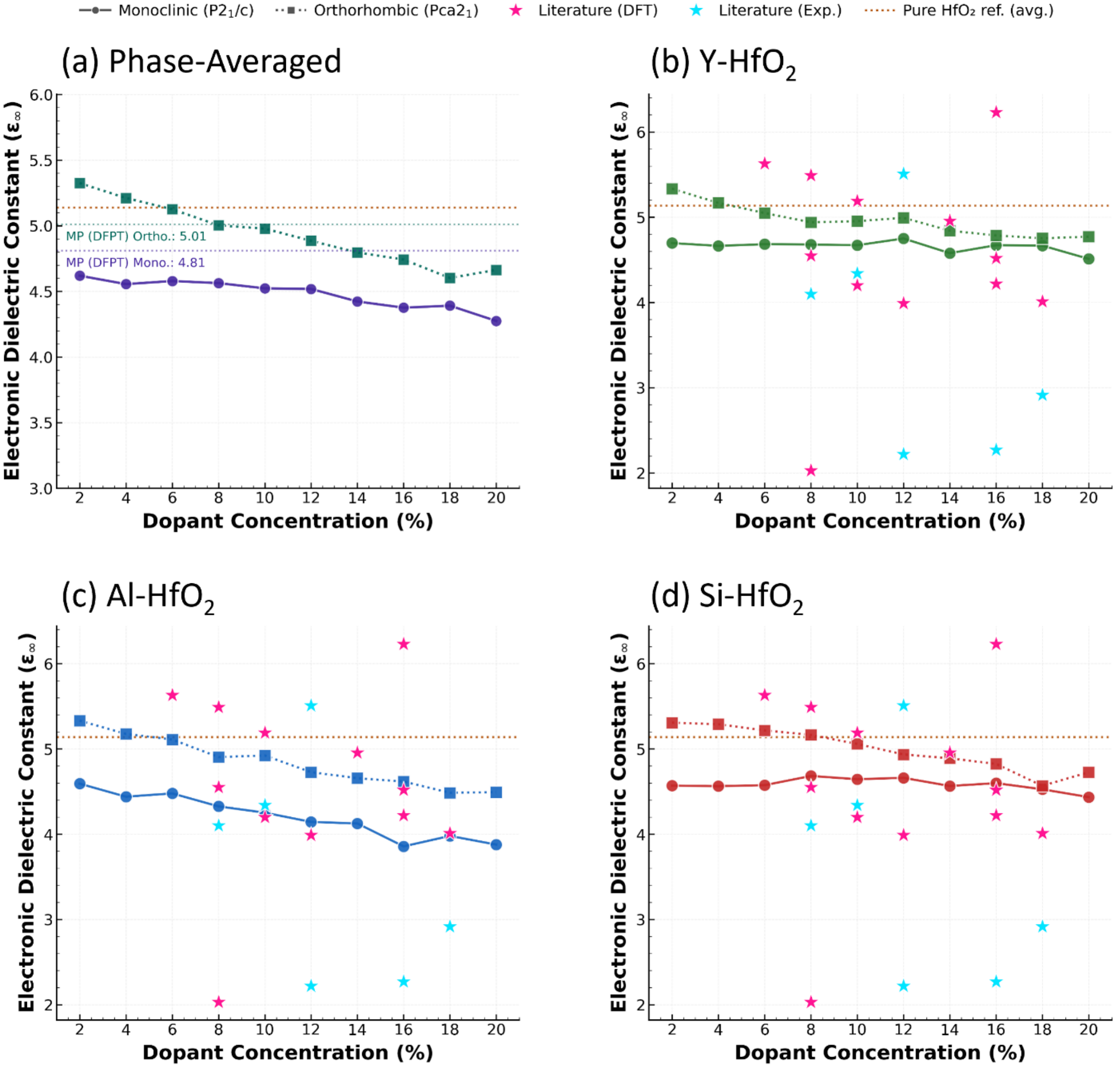

**Figure 5.** Electronic dielectric constant (ε) engineering in doped $HfO_2$. (a) Phase-averaged ε across all dopant systems for monoclinic (solid lines) and orthorhombic (dashed lines) phases, benchmarked against Materials Project DFPT reference values (horizontal dashed lines). (b–d) Concentration- and phase-resolved ε for (b) Y-, (c) Al-, and (d) Si-doped $HfO_2$. ALIGNN predictions (lines with markers) are compared against DFT literature (pink stars) and experimental

literature (turquoise stars), with the pure $HfO_2$ SevenNet baseline shown as a dotted reference line. Across all dopant systems, both phase selection and compositional tuning demonstrate measurable modulation of the optical dielectric response, with orthorhombic configurations consistently yielding higher $\varepsilon$ values relative to their monoclinic counterparts.

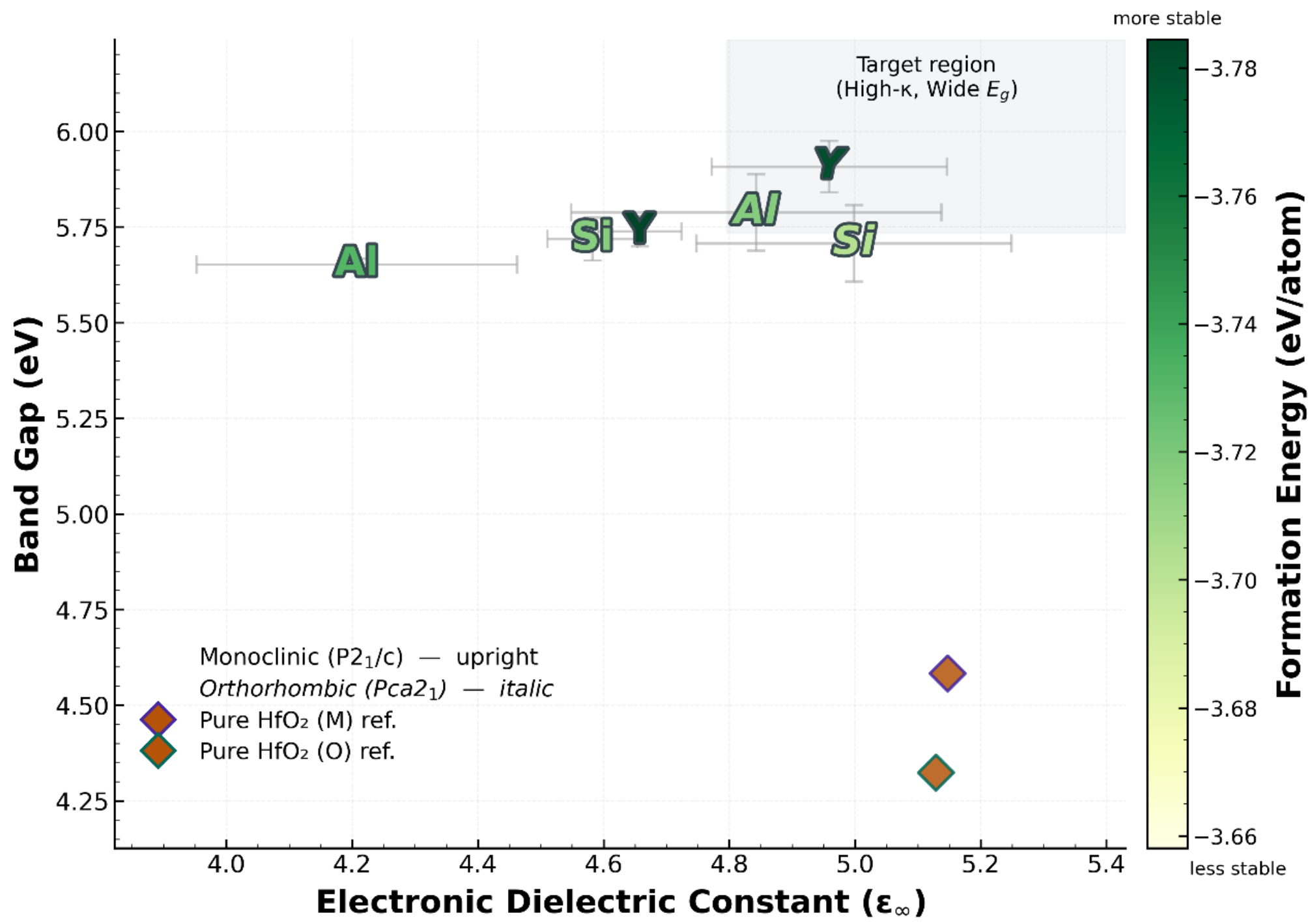

**Figure 6**. High-κ dielectric trade-off map for doped $HfO_2$. ALIGNN-predicted MBJ band gap ($E_g$) versus electronic dielectric constant (ε) for Al-, Si-, and Y-doped configurations. Data points are drawn as element symbols (Al, Si, Y) at concentration-averaged centroids, with upright type denoting the monoclinic phase and italic type the orthorhombic phase; error bars indicate ±1σ variance across the 2–20 at.% range. Symbol colour encodes the concentration-averaged formation energy (darker green = more negative $E_f$ = greater thermodynamic stability), allowing all three predicted properties to be read simultaneously. Dopant-induced displacements in the ε–$E_g$ space are benchmarked against pure $HfO_2$ references (diamonds, phase-coded edge colour). The shaded region denotes the optimal target space for gate dielectrics, requiring a simultaneous balance of high polarizability and a wide band gap to suppress leakage current.

| Target conc. (at.%) | Dopant atoms (of 108) | Realized conc. (at.%) | Remaining Hf sites |
|---|---|---|---|
| 2 | 2 | 1.9 | 106 |
| 4 | 4 | 3.7 | 104 |
| 6 | 6 | 5.6 | 102 |
| 8 | 9 | 8.3 | 99 |
| 10 | 11 | 10.2 | 97 |
| 12 | 13 | 12 | 95 |
| 14 | 15 | 13.9 | 93 |
| 16 | 17 | 15.7 | 91 |
| 18 | 19 | 17.6 | 89 |
| 20 | 22 | 20.4 | 86 |

**Table 1.** Mapping of target dopant concentration to the number of substituted Hf atoms across the 108 available cation sites.

## Supplementary Figures

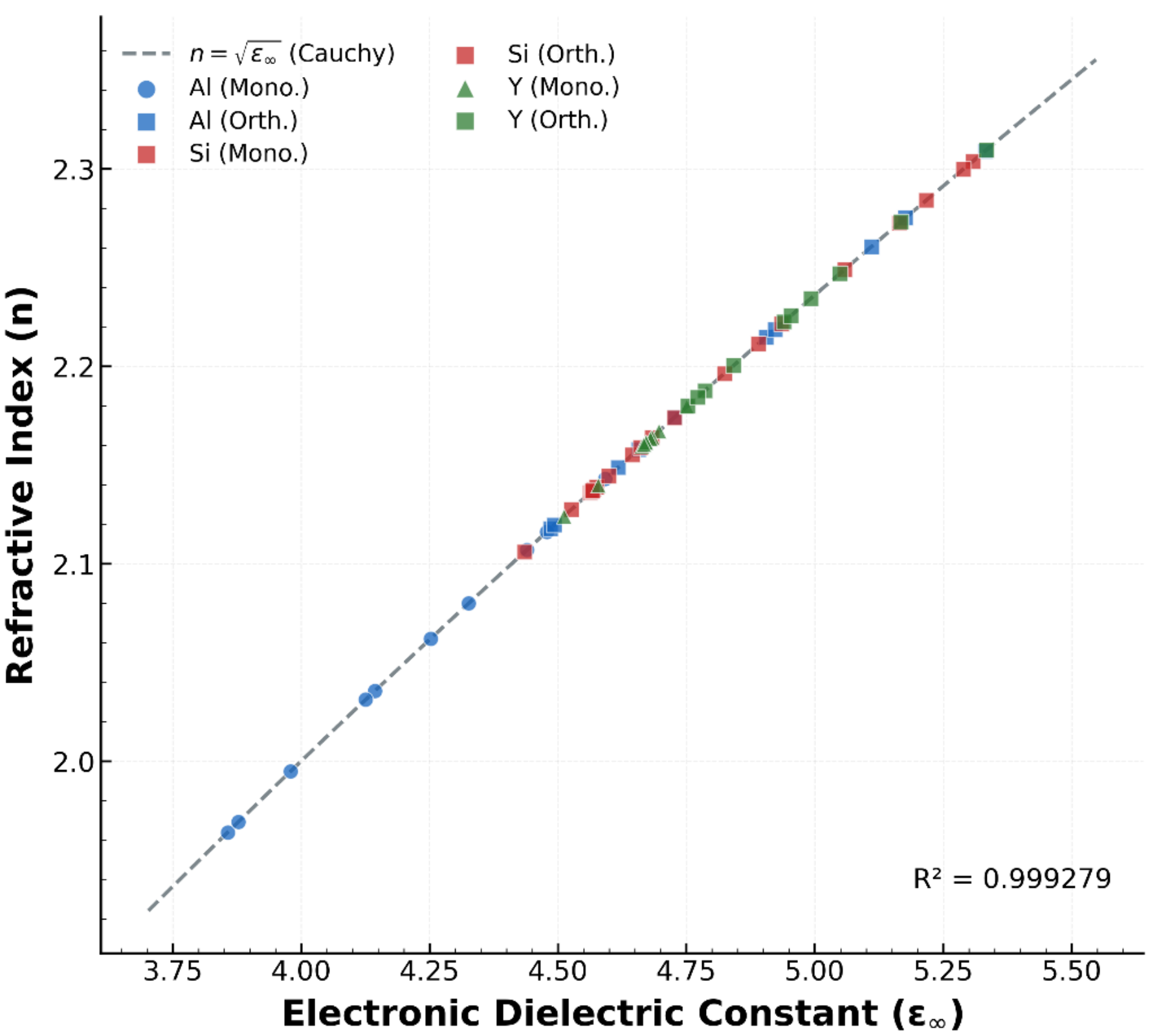

**Figure S1.** Optical dispersion correlation in doped $HfO_2$. Refractive index ($n$) versus electronic dielectric constant (ε) across all phase and dopant combinations. The dashed curve represents the ideal theoretical Maxwell relation ($n = \sqrt{\epsilon\infty}$). The near-unity coefficient of determination ($R^2$) confirms that the ALIGNN-predicted dielectric values maintain strict internal self-consistency with classical electromagnetic theory.

| **Model parameter** | **Value** |
|---|---|
| Host primitive cell | 12 atoms: 4 Hf + 8 O |
| Supercell expansion | 3 × 3 × 3 (27 primitive cells) |
| Total atoms per structure | 324 (108 Hf + 216 O) |
| Available Hf substitution sites | 108 |
| Concentration increment per dopant | ~ 0.93 at. % per Hf site |
| Force convergence criterion | $f_{max}$ = 0.01 eV/Å |
| Phase modeled | Monoclinic ($P2_1/c$), Orthorhombic ($Pca2_1$) |
| Dopants | Y, Al, Si |
| Total configurations | 60 (3 dopants × 2 phases × 10 concentrations) |

**Table S1.** Fixed supercell and computational parameters used in this study.